\documentclass[twocolumn,showpacs,preprintnumbers,amsmath,amssymb]{revtex4}
\usepackage{graphicx}
\usepackage{dcolumn}
\usepackage{bm}
\usepackage{amssymb}
\usepackage{amsmath}

\begin{document}
\title{Zero-bias anomalies on Sr$_{0.88}$La$_{0.12}$CuO$_2$ thin films}

\author{L. Fruchter, F. Bouquet and Z.Z. Li}%
\affiliation{Laboratoire de Physique des Solides, Univ. Paris-Sud, CNRS, UMR 8502, F-91405 Orsay Cedex, France}
\date{Received: date / Revised version: date}

\begin{abstract}{High-impedance contacts made on the surface of Sr$_{0.88}$La$_{0.12}$CuO$_2$ superconducting thin films systematically display a zero-bias anomaly. We consider two-level systems (TLS) as the origin of this anomaly. We observe that the contribution of some TLS to the contact resistance is weakened by a magnetic field. We show that this could result from the increase of the TLS relaxation rate in the superconducting state, due to its ability to create pairs of quasiparticles out of the condensate, when located close to the surface of the film.}
\end{abstract}


\maketitle

Mesoscopic contacts, smaller than the electron mean free path, are highly sensitive to both interactions between the material excitations and the conduction electrons, and to their scattering by impurities or defects present in the contact. Zero-bias anomalies, that are encountered in point-contact spectroscopy, are believed to often originate from the latter. As they may occur due to the external contamination of the contact, they are frequently not considered and kept out of spectroscopic studies, and thus remain poorly understood. However, they may also originate from defects that are characteristic of the bulk or the surface of the material investigated by the point-contact technique, and provide a unique way to access the intimate nature of atomic disorder in some cases. They were observed in contacts between clean simple metals\cite{Ralph1992,Keijsers1995}, and attributed in this case to scattering by two-level systems (TLS).

A TLS is encountered whenever a defect in a crystalline solid exhibits two metastable configurations. Such configurations result from the tunneling of a single atom, or a group of atom, between two positions with a small energy difference, separated by a large energy barrier, as compared to $k_B\,T$\cite{Black1981}. They are the typical excitations of glasses. While TLS in insulating glasses manifest themselves essentially by their interaction with phonons, in metallic glasses the coupling to conduction electrons is at the origin of their much faster relaxation than in dielectrics\cite{Doussineau1978,Golding1978}, as was first modeled in Ref.~\cite{Golding1978} (for a review, see Ref.~\onlinecite{Black1981}). As pointed out by Phillips\cite{Philipps1978}, glassiness is however not required and TLS may result from some short-range disorder.

Special attention has been paid to TLS in superconductors. Predictions for the interactions of TLS with the superconducting condensate were first made by Black and Fulde\cite{Black1979}. In particular, the influence of superconductivity on ultrasonic attenuation was considered. One of the predictions -- the decrease of the attenuation in the superconducting state, due to the decrease of the TLS relaxation rate as the interaction with the conduction electron is suppressed -- was indeed verified for metallic glasses \cite{Weiss1980,Arnold1981}. However, the prediction that TLS with a large splitting energy should find an additional channel for relaxation still lacks an experimental verification. Here, we show that TLS at the surface of a superconductor might be appropriate for the observation of such an effect.

We studied point-contact made at the surface of thin films of the electron-doped ``infinite layer'' Sr$_{1-x}$La$_{x}$CuO$_2$ compound (SLCO). These films grow epitaxialy with the (001) direction normal to the film. To elaborate point-contacts, we used 50 nm thick films grown by rf-magnetron sputtering on a (100) KTaO$_3$ substrate\cite{Li2009}. After growing the epitaxial film, a soft SLCO amorphous layer, about 10 nm thick, was deposited \textit{in situ} at room temperature. This layer protects the films from degradation, which is otherwise observed after a few weeks in air. It also provides an electrical insulating layer above the film which we used to elaborate contacts.

High-impedance contacts were made in two different ways. The first, most direct way, was to scratch the soft protection layer at the surface of the film with a tungsten tip, and press it mechanically. Such a method invariably yielded high-impedance contacts, with resistance from a few hundred ohms to a few k$\Omega$. For the second method, a gold layer was evaporated on the film, then patterned, leaving a checkerboard of dots about 250 $\mu$m large. By applying the tip of an ultrasonic bonding machine on the dots prior to the measurement, or by pressing the tungsten tip on the gold contact and vibrating it \textit{in situ}, holes could be created in the soft amorphous layer and the contact resistance could be tuned down from insulating to a few ohms. Although the resistance of such contacts could be found to increase upon aging at room temperature or thermal cycling, they were stable at low temperature and allowed for the study of the contacts over a large temperature interval. The spectroscopy of the low-impedance contacts provided insight into the superconducting order parameter that will be discussed elsewhere. High-impedance contacts made by the two above methods yielded similar results, and we present in the following data obtained from gold dots only.

The contact resistance was measured using two additional contacts deposited on the film, in a geometry minimizing the contribution from the film resistivity, when it is in the normal state. However, we could not avoid at high temperature a current-dependent contribution to the contact resistance, likely originating from the film. The differential resistance was measured using standard lock-in techniques, using an AC excitation current, in order to avoid measuring the macroscopic contact non-linearity, and a point-contact voltage inducing a broadening well below the thermal one, $V_{ac} \ll\,$k$_BT/e$.

\begin{figure}
\includegraphics[width= \columnwidth]{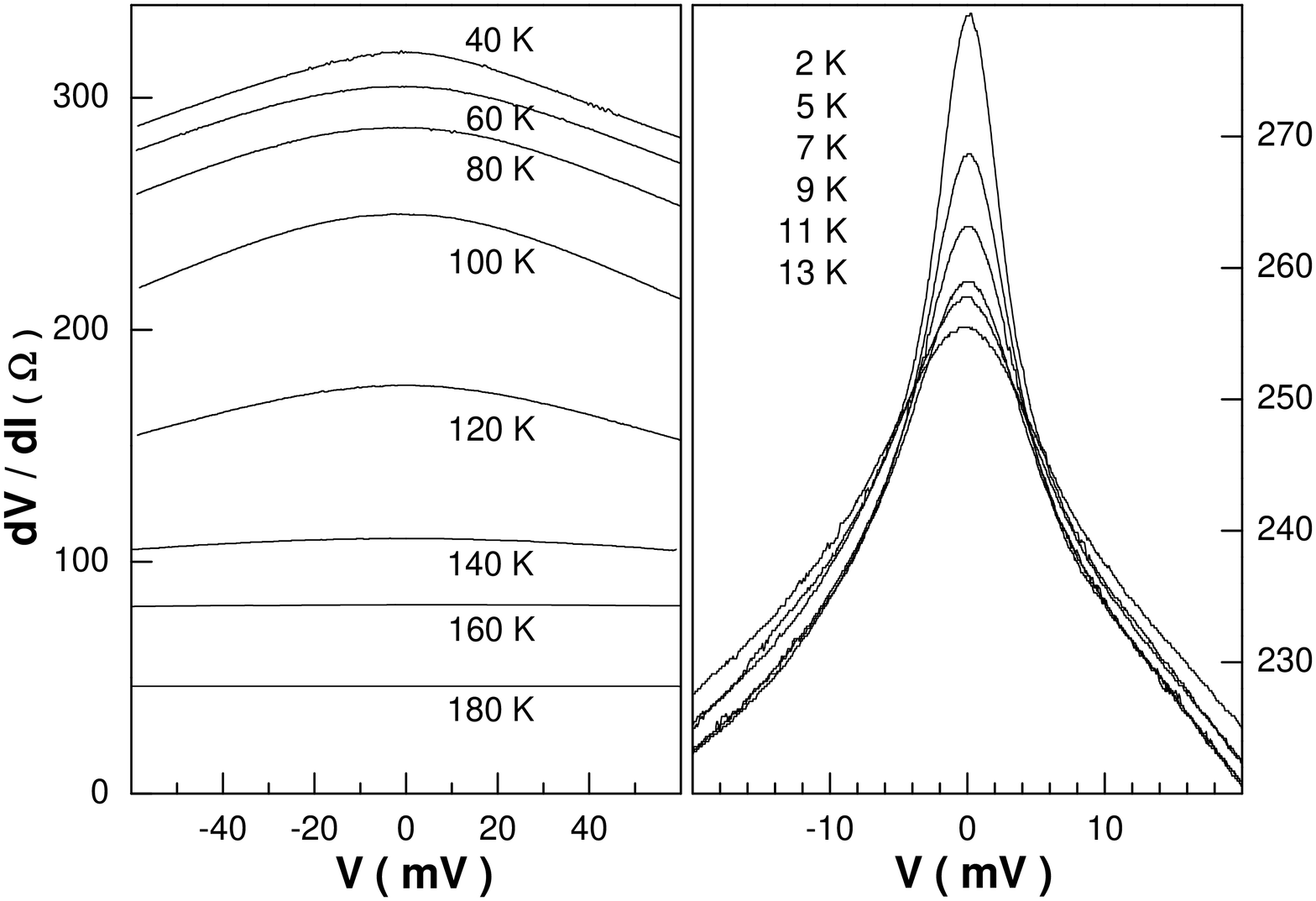}
\caption{Differential resistance for a high-impedance contact, showing a zero-bias anomaly developing at low temperature.}\label{rhighraw}
\end{figure}

\begin{figure}
\includegraphics[width= \columnwidth]{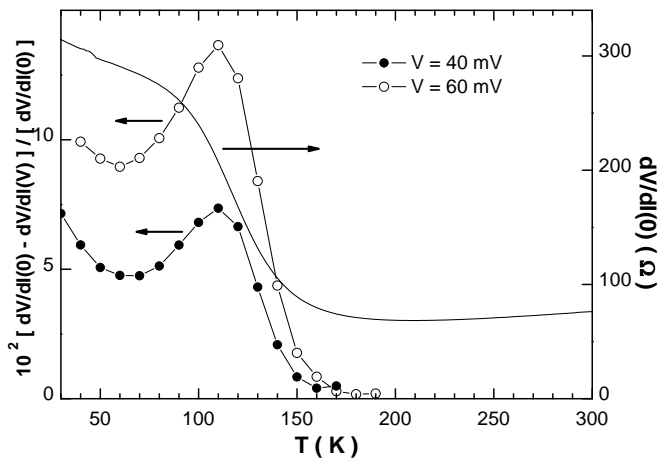}
\caption{Differential resistance and its bias voltage dependence for the high-impedance contact of Fig.~\ref{rhighraw}.}\label{rhigh}
\end{figure}

Figure~\ref{rhighraw} and~\ref{rhigh} display typical results obtained with a large resistance contact from a gold dot made on a film with $T_c =$ 20 K. Systematically, below $T \approx$ 150 K, the contact resistance increases strongly and develops a strong non-linearity with voltage bias. At low temperature, the smooth high-temperature non-linearity gives way to a sharp zero-bias anomaly, corresponding to a maximum of the differential resistance at $V = 0$. These data are representative of several contacts made at the surface of several films. Besides TLS, scattering by paramagnetic impurities is also known to yield zero-bias anomalies (for a review, see Ref.~\onlinecite{Yanson2005}). However, the Zeeman splitting in a magnetic field of the degenerate spin states leads to a characteristic splitting of the zero-bias anomaly into two peaks, and a dip at zero-bias in the contact resistance, with width about $2\,\textnormal g\,\mu_B\,H$, in this scenario\cite{Duif1987,Omelyanchouk1985}. No such features were observed in a 9 T magnetic field, and therefore we conclude that such a Kondo scattering is not at the origin of the observed anomalies. In the following, we show that scattering by TLS qualitatively and quantitatively account for our data.

First it is necessary to evaluate the regime applicable to our contacts. Point-contact spectroscopy requires that the contact radius, $a$, is less than the mean free path, $l$, in the investigated metal. The contact radius may be evaluated from the Sharvin formula for the contact resistance\cite{Sharvin1965}

\begin{equation}
R_S = \frac{4\rho l}{3\pi a^2}
\label{sharvin}
\end{equation}

which, using the fact that the product $\rho l$ is a constant in the Drude approximation, is equivalent to

\begin{equation}
R_S = \frac{4\pi^3}{S_c S_F} (\frac{e^2}{\hbar})
\label{kulik}
\end{equation}

where $S_c= \pi a^2$ is the contact surface and $S_F= \pi k_F^2$ is the extremal cross-section of the Fermi surface\cite{Kulik1977}. Like all other cuprates, SLCO is fairly anisotropic and the contact resistance should be dominated by the two-dimensional Fermi surface. Additional complication is expected in this case, as both an electron and a hole pocket contribute to the conductivity. Nevertheless, one can use the total Fermi surface for the evaluation of Eq.~\ref{kulik}. Using approximately 50\% of the first Brillouin zone, one obtains a typical value $a \simeq 400 (R_S\left[Ohm\right])^{-1/2}$ \AA. For the contact in Fig.~\ref{rhighraw}, this yields $a \simeq 25$ \AA. This should not be larger than the mean free path, as the coherence length inferred from upper critical field measurements is of the order of 40 \AA. Equivalently, Eq.~\ref{sharvin}, using $\rho \simeq$ 150 $\mu\Omega\, cm$\cite{Jovanovic2009b} yields $l \simeq$ 30 \AA. Thus, these rough estimates show that one can expect the contact dimension and the mean free path to be comparable. Therefore, there should be a diffusive correction to the Sharvin resistance (a fraction of the Maxwell resistance), from which a spectroscopy of the contact is actually possible.

\begin{figure}
\includegraphics[width= \columnwidth]{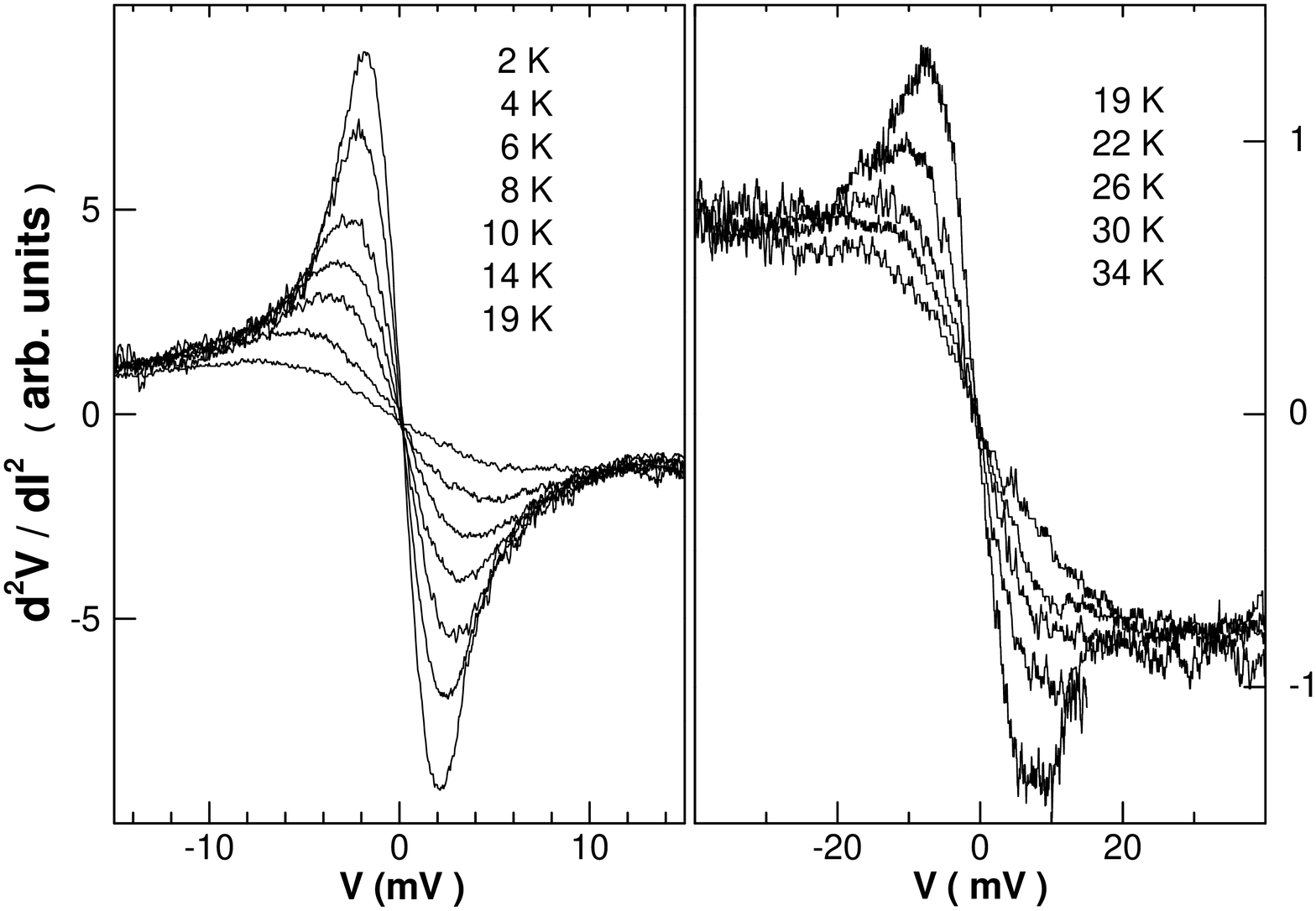}
\caption{Zero bias anomaly in $d^2V/dI^2$ for the high impedance contact in Fig.~\ref{rhighraw}.}\label{drdvhigh}
\end{figure}

\begin{figure}
\includegraphics[width= \columnwidth]{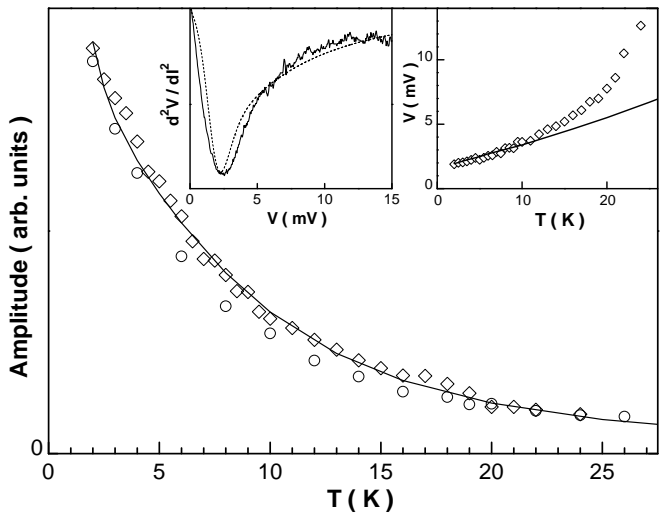}
\caption{Amplitude and bias voltage (right inset) for the peaks in Fig.~\ref{drdvhigh}. Diamonds are for $H = $ 0 T; circles are for $H = $ 9 T. Left inset shows the data at 4 K. Lines are fits using Eq.~\ref{kozubtls} with $q = $ 0.1 and $E =$ 1.7 meV.}\label{kozub}
\end{figure}

The scattering by TLS of electrons injected from point-contacts was first considered in Ref.~\cite{Kozub1984}. \textit{Inelastic} scattering by the TLS was found to result in a sharp peak in the point-contact spectrum, $d^2V/dI^2$, at $V=E$, where $E$ is the TLS energy splitting. The effect of the temperature is a finite peak width $\sim k_B\,T$. The data in Fig.~\ref{drdvhigh} clearly shows both a broadening and a shift to larger energy of the peak, as temperature increases, which rules out the inelastic scattering mechanism. It was pointed out that the \textit{elastic} scattering of the electrons by a slow TLS may dominate its resistance\cite{Kozub1986}. The TLS spectrum is determined in this case by the dependence of the two-level population on the energy of the incoming electrons, associated to two different cross sections for each state\cite{Akimenko1993,Keijsers1995}. The resistance for a collection of TLS with splitting energies $E_j$ is:

\begin{equation}
\frac{1}{R}\frac{dR}{dV} = \sum_j \frac{eC_j}{2E_j} (\sigma^{+}-\sigma^{-}_{j})\tanh(\frac{1}{2t})S(v,t,q)
\label{kozubtls}
\end{equation}

where the sum is over all TLS within the contact, $\sigma^{\pm}_{j}$ is the cross section for the upper and lower levels, $t=T/E_j$, $v=V/E_j$, $C_j$ is a constant, decreasing strongly with the TLS distance to the center of the contact, and $q$ is a geometrical factor varying between $q=1/2$ when the solid angle at which the contact is seen from the TLS is $2\pi$, and $q=0$ when the solid angle is zero. At $T=0$, $S(v,t,q)$ exhibits a $\delta$-function at $V=E_j$, which broadens to a strongly asymmetric peak, shifting to higher voltage as $T$ increases. Real contacts like the ones we have studied here are likely made of several mesoscopic contacts. We assume, however, that their resistance is dominated by a single low-resistance contact, with a unique TLS with energy splitting $E$ and geometrical factor $q$. The zero-bias anomaly may be``positive'' (showing a minimum in resistance at zero bias, and a positive peak in $d^2V/dI^2$ at positive bias) or``negative'', for respectively $\sigma^{+}_{j}-\sigma^{-}_{j}$ positive or negative. We always observed negative anomalies. 
In Fig.~\ref{kozub}, the amplitude and the position of the peak in $d^2V/dI^2$ may be tracked with temperature up to $T \simeq$ 25 K. For higher temperature, although some characteristic energy may be defined, $d^2V/dI^2$ saturates above this energy to a plateau (Fig.~\ref{drdvhigh}). As may be seen in Fig.~\ref{kozub}, the amplitude, the bias voltage and the profile of the peak may be reasonably accounted by Eq.~\ref{kozubtls}. However the fit for the peak bias voltage fails at large temperature, as the peak vanishes. This is likely due to the contribution of the weak non-linear contact resistance, which systematically develops below $T \simeq$ 150 K, the origin of which is unknown. We are not, either, aware of any TLS spectroscopic study that could check the validity of Eq.~\ref{kozubtls} at temperature as large as $E_j/k_B$ (see e.g. Refs~\onlinecite{Akimenko1993} and \onlinecite{Keijsers1995}).

Although there seems to be a general agreement to attribute the origin of the zero-bias anomaly, in the absence of paramagnetic impurities, to TLS, the elastic scattering model by Kozub and Kulik is actually challenged by the two-channel Kondo model\cite{Vladar1983}. For this model, a non-trivial $T=0$ fixed point governs the low temperature properties. It was found, however, that experimental data such as in Fig.~\ref{kozub} cannot easily decide between these two models\cite{Keijsers1995}. It has been proposed, instead, that the evaluation of the TLS relaxation rate can tell which of the two mechanisms is pertinent\cite{Balkashin1998}. 

What is, then, the effect of a magnetic field on the contact spectrum? In the two-channel Kondo model, a magnetic field breaks the spin degeneracy, yielding a linear decrease of the contact resistance at zero bias with applied magnetic field\cite{Ralph1992,Ralph1994}. It is however difficult to go beyond the scaling approach to evaluate the magnitude of the effect. In Kozub and Kulik's model, no effect of the magnetic field is expected when the TLS results from two structural defect configurations. It was, however, pointed out that TLS may also result from electronic disorder \cite{Kozub1996}. In this case, a magnetic field, which reduces the amplitude of mesoscopic fluctuations, will also reduce the magnitude of the zero-bias anomaly, as for the Kondo model.

\begin{figure}
\includegraphics[width= \columnwidth]{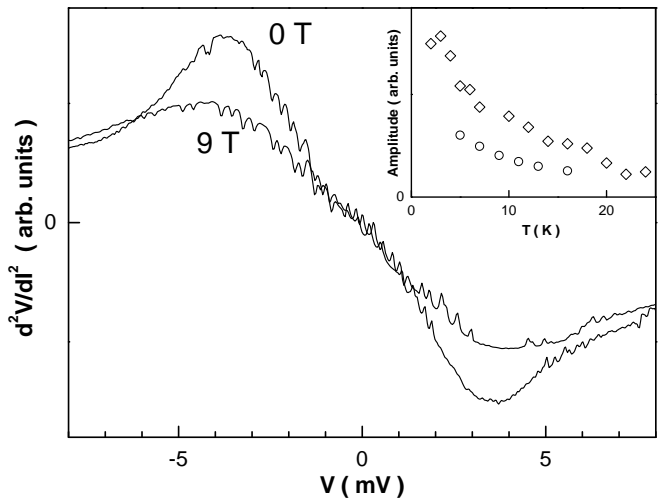}
\caption{Effect of the magnetic field on a high impedance contact spectrum ($T = 7$ K). The inset displays the amplitude of the peak in $d^2V/dI^2$. Diamonds are for $H$ = 0 and circles for $H$ = 9 T.}\label{field}
\end{figure}

In our case, we observe that the effect of the magnetic field on the zero-bias anomaly varies greatly from contact to contact. In Fig.~\ref{kozub}, only a slight decrease of the amplitude anomaly could be observed upon applying the magnetic field. For some contacts, the magnetic field is able to depress the amplitude much more strongly, as can be seen in Fig.~\ref{field}. Within the Kondo model, it is difficult to explain such a dispersion: the effect of the magnetic field is intrinsic to the scattering mechanism, so we expect comparable magnitudes of the effect for contacts showing comparable spectra, as observed in Ref.~\onlinecite{Ralph1994}. In the case of mesoscopic electronic disorder, the magnetic field should either show variable fingerprints, with random sign -- which is not observed -- or an average effect, resulting from the summation over many fluctuators\cite{Kozub1996}. As for the Kondo model, we expect for the latter case comparable magnitudes for comparable spectra. Therefore, we consider another mechanism for the effect of the magnetic field, taking into account the interaction of the TLS with superconductivity.

Both samples in Fig.~\ref{kozub} and \ref{field} showed bulk superconductivity at $T_c \approx$ 20 K. The applied magnetic field, 9 T, drives the sample to its normal state over a large temperature range\cite{Jovanovic2009}. As the magnetic field destroys superconductivity, TLS that are strongly coupled to the superconducting condensate can either gain or loose a channel to relax. Indeed, it is known that the enhanced TLS relaxation in metals is analogous to the Korringa relaxation of nuclear spins interacting with conduction electrons, and the relaxation rate depends on the square of the density of electronic states at the Fermi energy\cite{Golding1978}. As a consequence, the superconducting gap eliminates this relaxation channel for the TLS, as long as the TLS energy, $E$, is smaller than the gap\cite{Black1979}. However, when $E > 2\Delta$, the relaxation may occur by the creation of a pair of quasi-particules. The relaxation rate being the one for a time-reversal-invariant interaction of a pseudospin with the condensate\cite{Black1979}, the coherence factor is of the order of unity\cite{Tinkham1975} and the relaxation rate is \textit{larger} in the superconducting state than in the normal state, up to a factor $\sim 2$.
The first mechanism is at the origin of the variation in the superconducting state of the ultrasonic absorption in amorphous superconductors\cite{Thomas1980,Arnold1981}. However, to the best of our knowledge, there has been to date no observation by ultrasonic attenuation or velocity measurements of the second mechanism. Presumably, this is due to the fact that it may concern glassy superconductors that are so disordered, that they actually are gapless or no longer superconducting. The situation may be more favorable when disorder is probed at the surface of a superconductor, as in the present case.

As shown in Refs.~\onlinecite{Kozub1984} and \onlinecite{Kozub1986}, the relative intensities of the elastic and inelastic contributions of the TLS to non-linearity are governed by the TLS relaxation rate, the former being favored in the case of slowly relaxing systems. We then expect that an increase of the relaxation rate, as would be the case for the quasiparticules-pair generation mechanism, should favor the latter. The enhanced inelastic scattering in the superconducting state should be accompanied by a shift to lower energy and a narrowing of the zero-bias peak. This is indeed what we observe (Fig.~\ref{field}). Within such a picture, the TLS in Fig.~\ref{kozub} is only weakly coupled to the condensate, whereas the behavior in Fig.~\ref{field} is one of a strongly coupled TLS.

The requirement $E > 2\Delta(T)$ for the pair generation is however a strong constraint. Using $\Delta(0)/\textnormal k_\textnormal B T_c \simeq 3.5$, as found for optimally doped SLCO\cite{Chen2002,Khasanov2008}, requires $E \gtrsim$ 6 meV for $T_c$ = 20 K, which is quite a large value. The superconducting gap may however be locally smaller, considering that the TLS may be present in some part of the film which is strongly disordered, or close enough to the interface so as to experience a depressed gap. There is also a requirement on the location of the TLS, which should be located within the superconductor's coherence length from the interface: further away into the superconductor, the point contact resistance would be set by the Andreev reflection mechanism, whereas it can experience both the quasi-particles from the normal metal and the superconducting condensate close to the interface. We note that some point-contacts have been observed, between a metal and a superconductor, exhibiting both Andreev reflection and noise from TLS, as well as a large broadening parameter, suggesting pair-breaking effects\cite{Wei2010}.

It should be possible to bring such a mechanism into play for conventional superconductors, and observe a direct TLS signature as well as their interaction with superconductivity. We would like to suggest some ways to perform this. What one needs is actually a high density of TLS segregated in a metal, subject to the proximity effect from a superconductor. This may be, as in the present case, defects at the surface of a superconductor, or a sandwich made of a superconductor covered by a thin glassy metal layer, both systems probed using point-contact spectroscopy. Alternatively, one could think of a short normal metal wire with a TLS, coupled to a superconducting bank, although we are not aware of any TLS created in a controlled way in a metal wire. When using a glassy metal, it should be kept in mind that the contact dimension should be as small as the reduced mean free path in the disordered material, in order to allow for spectroscopy: this suggests using techniques able to finely scan surfaces, such as AFM in contact mode.

\begin{acknowledgments}
The authors are grateful to G. Collin for valuable contributions on materials elaboration and structural studies.
\end{acknowledgments}

\newpage

\end{document}